\documentclass[twocolumn,numberedappendix,appendixfloats,apjl,tighten]{openjournal}

\usepackage{xcolor}
\usepackage{textgreek}
\usepackage[utf8]{inputenc}
\usepackage[english]{babel}

\usepackage{hyperref}
\hypersetup{
    unicode, 
    colorlinks=true,
    linkcolor=linkcolor,
    citecolor=linkcolor,
    filecolor=linkcolor,
    urlcolor=linkcolor,
}
\usepackage{color,colortbl}
\definecolor{linkcolor}{rgb}{0.0,0.3,0.5}

\newcommand{\rev}[1]{{#1}}
\usepackage{tensind}
\tensordelimiter{?}
\DeclareGraphicsExtensions{.bmp,.png,.jpg,.pdf}
\usepackage{verbatim}
\usepackage[normalem]{ulem}
\usepackage{orcidlink}
\usepackage{soul}

\usepackage{gensymb}

\def\frontmatter@above@affilgroup{\vspace*{2pt}}

\begin{document}
\title{Exploring TRAPPIST-1 Climate States with an Energy Balance Model\vspace{-1.5cm}}

\author{Jacob Haqq-Misra\orcidlink{0000-0003-4346-2611}}
\email{jacob@bmsis.org}
\affiliation{Blue Marble Space, 600 1st Avenue, 1st Floor, Seattle, Washington 98104, USA}

\begin{abstract}
This paper presents a version of the HEXTOR energy balance model that has been configured for the study of habitable terrestrial planets orbiting low-mass stars. The model is validated for rapidly-rotating Earth-like planets using latitudinal coordinates, which shows expected patterns of bistability. A tidally-locked coordinate transformation is then applied to the model, which is calibrated to match mean values of the minimum, average, and maximum surface temperatures from a general circulation model ensemble of TRAPPIST-1 e. This calibrated energy balance model is used to characterize the possible climate states of such a synchronously rotating planet across a parameter space of instellation and carbon dioxide partial pressure. These calculations suggest a state of partial ice cover for TRAPPIST-1 e and complete ice cover for TRAPPIST-1 f. \rev{TRAPPIST-1 e becomes fully ice-free only above $\sim$0.4 bar CO$_2$, while TRAPPIST-1 f remains ice-covered unless CO$_2$ partial pressure approaches $\sim$1.2 bar.} This approach demonstrates the capability of a simplified one-dimensional model to study the climates of terrestrial planets in synchronous rotation, which can help guide more complex models and observations toward the most promising targets of interest.
\end{abstract}

\begin{keywords}
    {astrobiology -- exoplanet atmospheres -- M dwarf stars -- habitable planets}
\end{keywords}

\maketitle

\section{Introduction}

Terrestrial planets orbiting low-mass stars remain optimal candidates for discovery and characterization through transit observations (see, e.g., \cite{shields2016habitability} for a review). M-dwarf stars comprise about three-fourths of the total stellar population of the galaxy, and estimates based on \textit{Kepler} data suggest that at least half of such systems should contain terrestrial planets within the liquid water habitable zone \citep[e.g.,][]{kopparapu2013revised}. Observations with \textit{JWST} have enabled atmospheric characterization of such systems for the first time, primarily focused on the TRAPPIST-1 planets \citep[e.g.,][]{lincowski2023potential,ducrot2025combined}, and continued transit observations will enable more precise constraints on the presence and composition of any atmospheres \citep{de2024roadmap}.

Planning and interpreting observations includes the use of computational climate models, which provide constraints on physically plausible atmospheric configurations and the observational strategies required to characterize them \citep[e.g.,][]{lustig2019detectability}. M-dwarf planets within the liquid water habitable zone may fall into synchronous rotation with their host star, which leads to an atmospheric configuration with one side in perpetual daylight (the dayside or sub-stellar hemisphere) and the other side in perpetual night (the nightside or anti-stellar hemisphere), divided by the terminator in perpetual twilight. Three-dimensional general circulation climate models have shown that a wide variety of such atmospheres are stable, avoiding collapse from the atmosphere freezing out on the night side due to sufficient energy transport by atmospheric dynamics \citep[e.g.,][]{joshi1997simulations,joshi2003climate,merlis2010atmospheric,edson2011atmospheric,carone2014connecting,kumar2016inner,turbet2016habitability,del2019habitable,lefevre20213d}. Other recent studies have used general circulation models to explore variations in atmospheric properties and construct simulated observed spectra that can provide a basis for interpreting transit observations \citep[e.g.,][]{kiang2021land,mak20243d,adams2025habitability,wolf2025chemistry}. 

General circulation models are complex and computationally expensive, which include explicit calculation of radiative transfer, atmospheric dynamics, and other physical processes. Such models can be useful for studying important problems for Earth and other planets, but the ability to explore large parameter spaces with general circulation models remains computationally challenging. Some studies have performed sparse sampling experiments that attempt to maximize the amount of information gained from a multi-dimensional parameter space while minimizing the number of simulations \citep{kiang2021land,adams2025habitability,wolf2025chemistry}. Another complementary approach is to leverage simpler and more computationally efficient models to explore large parameter spaces; such results can be instructive by themselves and can also provide guidance for the regions of parameter space to study in detail with more complex models.

This study presents an updated version of HEXTOR (Habitable Energy balance model for eXoplaneT ObseRvations) for simulating climate states of a TRAPPIST-1 planet in synchronous rotation. Energy balance models (EBMs) are one-dimensional latitudinal models that have traditionally been applied to study climate bistability on Earth \cite[e.g.,][]{north2017energy}, and various EBMs have also been used to study rapidly-rotating (i.e., Earth-like) climates with application to exoplanets \citep[e.g.,][]{deitrick2023functionality,barnes2025functionality}. The previous version of HEXTOR \citep{haqq2022energy} implemented a tidally locked coordinate system, which was able to demonstrate a qualitative application of using an EBM to study M-dwarf planetary atmospheres. The version of HEXTOR in this study has been calibrated against results from a general circulation model ensemble of TRAPPIST-1 e \citep{sergeev2022trappist}, which represents a first demonstration of the capability of a one-dimensional model to provide quantitative insight on M-dwarf climates.

The paper begins by tuning the updated model to pre-industrial Earth conditions, exploring sensitivity to obliquity, and showing the resulting bistable climates for Earth-like planets. The following section utilizes a tidally-locked coordinate transformation to simulate a reference TRAPPIST-1 e case with 1\,bar N$_2$ and 400\,ppm CO$_2$, calibrated to the results of a general circulation model ensemble. This calibrated model is then used to explore a parameter space spanning instellation and CO$_2$ partial pressure, which can provide constraints that apply to the characterization of TRAPPIST-1 e and f. 

\vspace{0.5cm}\section{Earth Tuning}

\rev{This section validates the updated model for rapidly-rotating Earth-like planets, first by tuning to the pre-industrial Earth climate and then by comparing the resulting bistability structure against published EBM results, before the model is applied to synchronous rotators in the following section.} The updated version of HEXTOR presented here uses the same one-dimensional energy balance equation as other Budyko-Sellers-type models \cite[e.g.,][]{north2017energy}:
\begin{equation}
    C\frac{\partial T}{\partial t}=S\left(1-\alpha\right)-F+\frac{\partial}{\partial x}\left[D\left(1-x^2\right)\frac{\partial T}{\partial x}\right].\label{eq:EBM}
\end{equation}
This equation calculates the change in surface temperature ($T$) with time ($t$) at each latitudinal band ($x$, defined as the sine of latitude). The first term on the right-hand side of Eq. (\ref{eq:EBM}) represents incoming stellar energy, with $S$ as the total instellation and $\alpha$ as the planetary albedo. The term $F$ represents outgoing infrared radiation. The last term in  Eq. (\ref{eq:EBM}) is a diffusive latitudinal energy transport, with $D$ representing thermal conductivity. The value of heat capacity ($C$) can vary with land fraction and ice coverage.

The core model construction and parameters are unchanged from the previous version of HEXTOR \citep{haqq2022energy}, which are briefly described here. The model solves Eq. (\ref{eq:EBM}) by using 18 latitudinal bands spaced ten degrees apart, with a $\Delta t=12$\,hr finite differencing time step. Each latitudinal band has a fixed land fraction and is assigned an initialization temperature. Seasonal variation in instellation is captured as a time-dependent function for $S$, with the mean value of $S$ specified at runtime. Heat capacity, $C$, is a weighted average based on surface coverage, with a value of $C_{\text{land}}=5.25\times10^6$\,J\,m$^{-2}$K$^{-1}$ for land, $C_{\text{ocean}}=40C_{\text{land}}$ for ocean, and $C_{\text{ice}}=2C_{\text{land}}$ for ice. The diffusive parameter is a constant 0.58\,W\,m$^{-2}$\,K$^{-1}$ (note that $D=0.38$\,W\,m$^{-2}$\,K$^{-1}$ was used by \cite{haqq2022energy}). The model calculates surface albedo at each latitudinal band (which will be used in the functional forms of $\alpha$ and $F$ discussed next). Surface albedo depends on surface coverage, with a value of 0.2 for unfrozen land, 0.663 for ice, and a zenith-angle dependent value for unfrozen ocean \citep{williams1997habitable}.

The values of planetary albedo and outgoing infrared radiation are obtained by interpolating data in a lookup table \citep{haqq2022energy}. The HEXTOR lookup table contains values of $F$ and $\alpha$ that were calculated with a radiative-convective equilibrium climate model \citep{kopparapu2013habitable}, assuming a fixed 1\,bar N$_2$ atmosphere with variable CO$_2$ \rev{(see Appendix \ref{ap:updates} for the parameter ranges)}. The advantage of this approach is to allow $\alpha$ and $F$ to vary with changes to atmospheric composition, in a way that remains consistent with theoretical expectations from existing validated models. \rev{Within the lookup table, surface albedo is an input parameter, so the presence of surface ice is imposed by the EBM rather than predicted by the radiative-convective model. At fixed surface albedo, the variation of $\alpha$ with CO$_2$ reflects gaseous radiative transfer alone: for a solar spectrum, planetary albedo increases steeply with CO$_2$ partial pressure above ${\sim}0.2$\,bar due to Rayleigh scattering, whereas for a 2600\,K M-dwarf spectrum (used in the following section) planetary albedo instead decreases with increasing CO$_2$, because near-infrared absorption by CO$_2$ and H$_2$O outweighs the weak Rayleigh scattering at the wavelengths where such stars emit.} The radiative-convective model used to generate the lookup table data is capable of calculating upward- and downward-directed radiative transfer in an atmospheric column to find a steady-state vertical temperature profile; however, such models cannot represent horizontal spatial variation. The approach of using a lookup table, or a high-order polynomial fit \cite[e.g.,][]{williams1997habitable,haqq2016limit}, to parameterize these radiative transfer variables can allow for an EBM to explore parameter spaces that would be challenging with more computationally expensive models. This present version of HEXTOR includes corrections to the radiative transfer implementation, which are discussed in Appendix \ref{ap:updates}. 

\begin{figure*}[tbp]
    \centering
    \includegraphics[width=\textwidth]{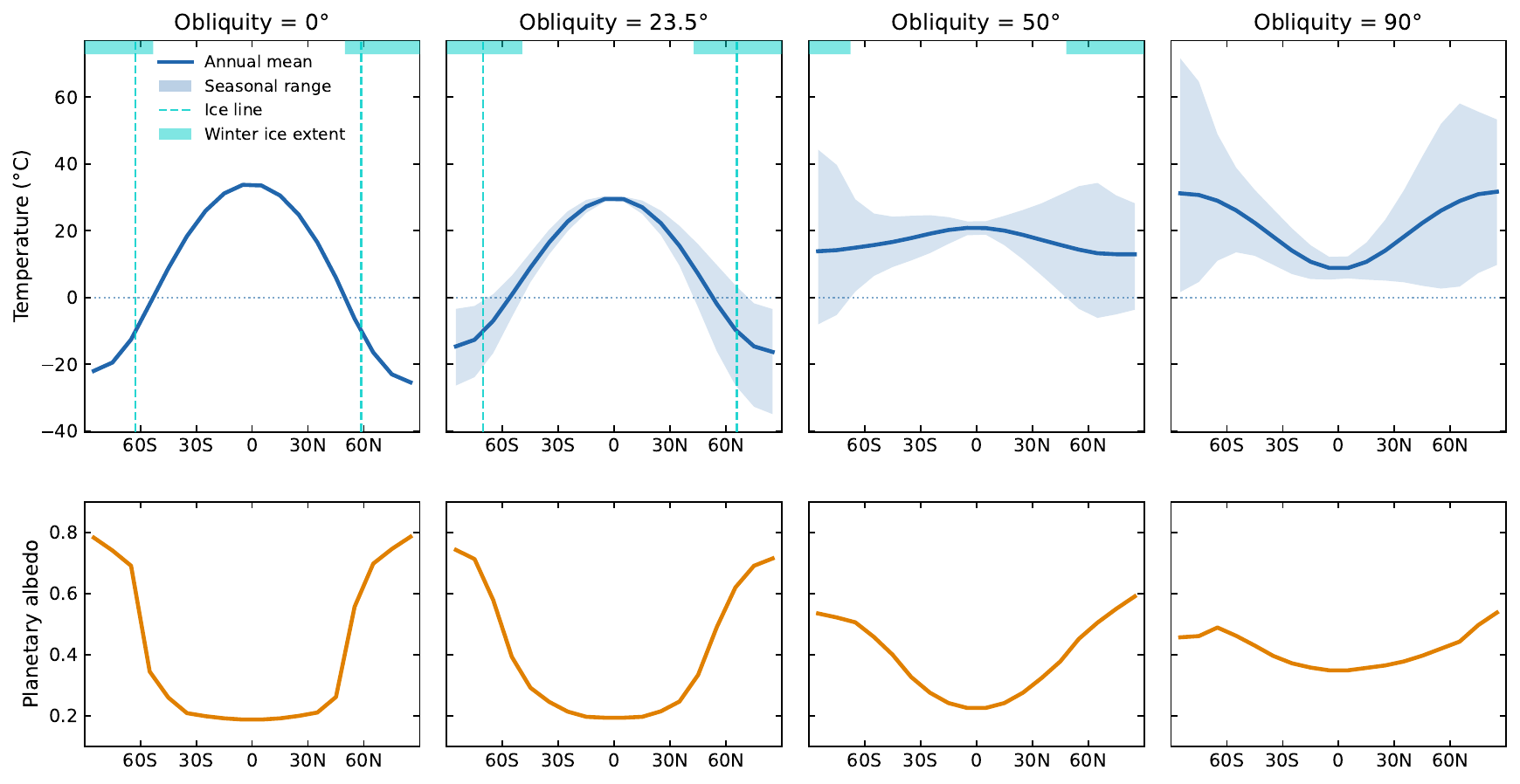}
    \caption{Calculations with the improved HEXTOR model of average temperature with seasonal range (top row) and planetary albedo (bottom row) for an Earth-sized planet orbiting the Sun with 1 bar N$_2$ and 280 ppm CO$_2$, with obliquity ranging from 0$\degree$ to 90$\degree$. Calculations assume a present-day Earth land fraction at each latitudinal band. \rev{Dashed vertical lines show ice lines, defined where the annual-mean temperature crosses the 263.15\,K ice threshold (i.e., permanent ice). Shaded bands along the top of each panel show the winter ice extent, defined where the seasonal minimum temperature falls below freezing.}\vspace{0.20cm}}
    \label{fig:Earth}
\end{figure*}

\begin{figure*}[tbp]
    \centering
    \includegraphics[width=\textwidth]{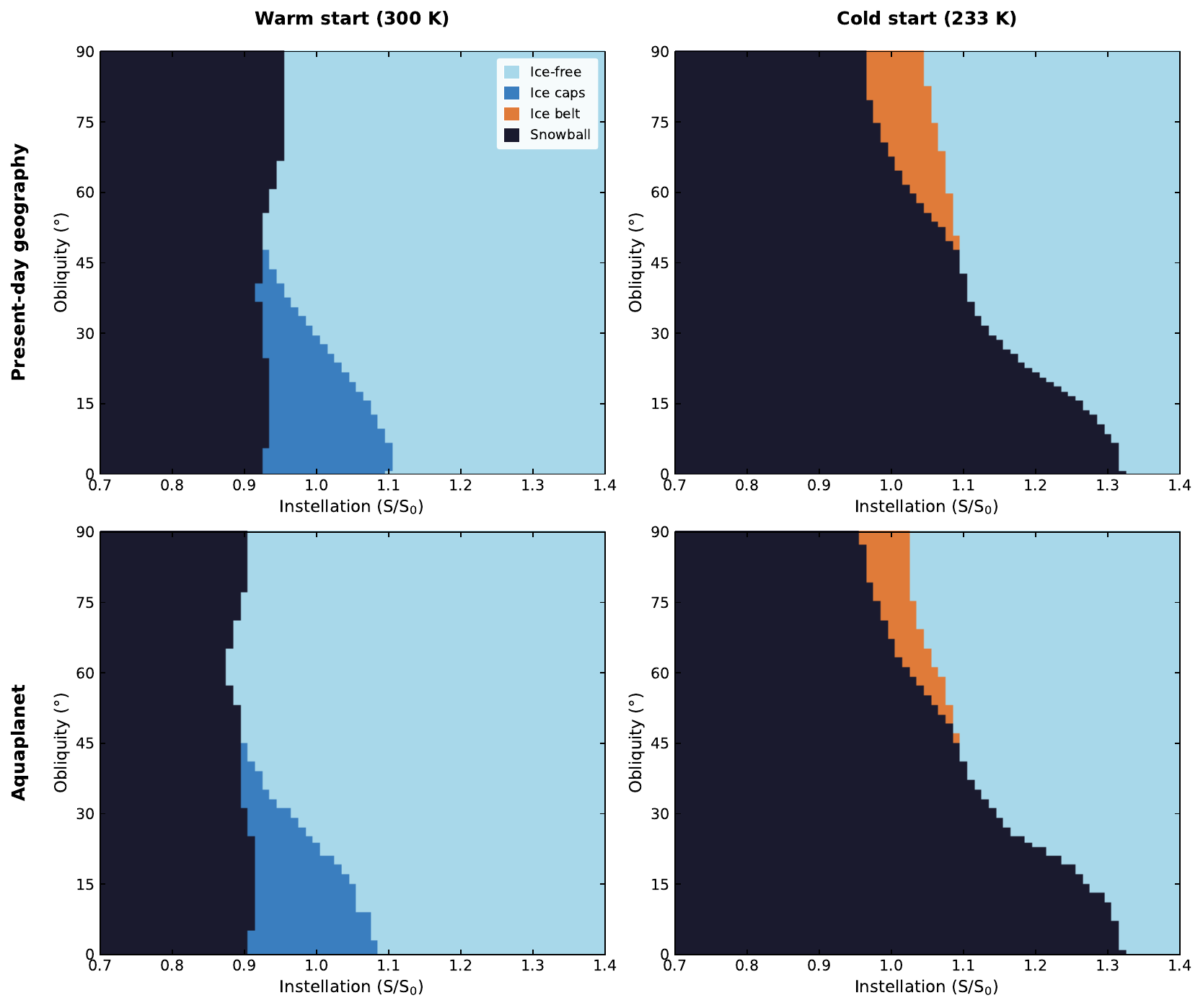}
    \caption{Ice states with the improved HEXTOR model as a function of instellation ($S/S_0$) and obliquity, for a warm start (left column) and a cold start (right column). The top row shows an Earth-sized planet with present-day geography, as in the calculations shown in Figure \ref{fig:Earth}. \rev{The bottom row shows an aquaplanet (no land) configuration, which allows direct comparison with the analytic aquaplanet EBM of \cite{rose2017ice}: stable ice caps occur only at obliquities below ${\sim}45$\degree, while ice belts occur only at higher obliquities and are only reached from cold initial conditions.} An ice belt state refers to conditions with ice-free poles and an equatorial glacial band that arise at high obliquity.\vspace{0.20cm}}
    \label{fig:phases}
    \label{fig:aqua}
\end{figure*}

The baseline pre-industrial Earth case reaches a steady-state radiative balance with an annual mean surface temperature of 288.0\,K, which assumes a 1\,bar N$_2$ atmosphere with an additional 280\,ppm CO$_2$. A summary of this climate state with variations in obliquity is shown in Figure \ref{fig:Earth}, with global average values listed in Table \ref{tab:means}. The hemispheric asymmetries result from the use of an Earth-like geography, represented as the land fraction at each latitudinal band \citep{williams1997habitable}. These solutions show a realistic seasonal range for a present-Earth obliquity, a lack of seasonality when obliquity is zero, and an inverted temperature profile at high obliquity. Ice lines are present for the 0\degree and 23.5\degree obliquity cases, but the 50\degree and 90\degree cases remain above freezing \rev{in the annual mean. Seasonal ice still occurs in the 50\degree\ case, where winter minimum temperatures drop below freezing poleward of $\pm$55\degree\ latitude and produce partial wintertime ice cover (reaching ${\sim}55$\% ice fraction at $\pm$85\degree) that melts each summer; in the 90\degree\ case, all latitudes remain above freezing throughout the year (Fig. \ref{fig:Earth})}.  
\begin{table}[t]
    \centering
    \begin{tabular}{cccc}
        \hline
        Obliquity & $T$ (K) & $\alpha$ & $F$ (W\,m$^{-2}$) \\
        \hline
        0\degree    & {288.8} & {0.251} & {255.2} \\
        23.5\degree & {288.0} & {0.252} & {253.2} \\
        50\degree   & {291.0} & {0.235} & {260.0} \\
        90\degree   & {290.6} & {0.234} & {260.0} \\
        \hline
    \end{tabular}
    \caption{Global average values calculated for Earth-like planets.}
    \label{tab:means}
\end{table}

As with other Budyko-Sellers-type models, HEXTOR calculations show hysteresis, with solutions that depend on the initialization choice for $T$. This leads to situations in which an Earth-like planet can reside in a stable configuration as either an ice-covered snowball, an ice-cap state like present Earth, or a completely ice-free state---with the same value of instellation. This phase space is plotted in the top row of Figure \ref{fig:phases}, which shows the resulting steady-state climate from initializing the model in a warm start at 300\,K (left column) compared to a cold start at 233\,K (right column). An additional ice belt state refers to conditions with ice-free poles and an equatorial glacial band that arise at high values of obliquity. This phase space is generally consistent with EBM results shown by \cite{wilhelm2022ice}, with differences that arise due to the use of a more complex representation of ice sheet flow (see also \cite{barnes2025functionality} for details about an ongoing EBM intercomparison project). Transitions between different climate states can be driven by changes in instellation (i.e., varying $S$) or by changes in the atmospheric greenhouse gas inventory (i.e., varying $F$).

\rev{A more quantitative validation of this bistability structure can be made against the annual-mean analytic aquaplanet EBM of \cite{rose2017ice}, whose Earth-like reference parameters (transport efficiency $\delta = 0.31$, albedo feedback $\alpha = 0.44$) closely match the equivalent HEXTOR values ($D/B \approx 0.29$ for an infrared sensitivity of $B \approx 2$\,W\,m$^{-2}$\,K$^{-1}$). The bottom row of Figure \ref{fig:aqua} shows the warm-start and cold-start phase space for HEXTOR in an aquaplanet configuration, computed over obliquities from 0\degree\ to 90\degree\ in steps of 2\degree. The regime structure agrees with the analytic model: stable polar ice caps occur at low obliquity (with an ice edge at 83\degree\ latitude for $S/S_0=1$ and 23.5\degree\ obliquity, compared to ${\sim}70$\degree\ for present Earth); the minimum instellation for an ice-free state decreases with obliquity, with a ratio between 90\degree\ and 23.5\degree\ obliquity of $0.91/1.01 = 0.90$ in HEXTOR compared to $1.06/1.2 \approx 0.88$ in the analytic model; and equatorial ice belts replace polar caps above a critical obliquity of ${\sim}45$\degree\ in HEXTOR, compared to the analytic value of 55\degree. The HEXTOR ice-cap state is somewhat more stable than the analytic prediction, with the snowball transition at 23.5\degree\ obliquity occurring at $S/S_0 = 0.91$ (compared to ${\sim}0.98$) and deglaciation at $S/S_0 = 1.20$ (compared to ${\sim}1.38$); such differences are expected from the seasonal cycle, the smooth fractional ice parameterization, and the nonlinear dependence of $F$ on temperature in HEXTOR, in contrast to the step-function albedo and linearized emission of the annual-mean analytic model. Note that ice belts in HEXTOR appear only on the cold-start branch, at obliquities of 46--90\degree\ and instellations near $S/S_0 \approx 1.0$--$1.1$, while warm starts at high obliquity produce only snowball or ice-free states: this directly illustrates the results of \cite{rose2017ice} that stable ice belts often coexist with both ice-free and snowball attractors and are inaccessible from most initial conditions, which may explain why ice belts are rarely found in three-dimensional simulations.}

\section{Synchronous Rotation}

\begin{figure*}[tbp!]
    \centering
    \includegraphics[width=\textwidth]{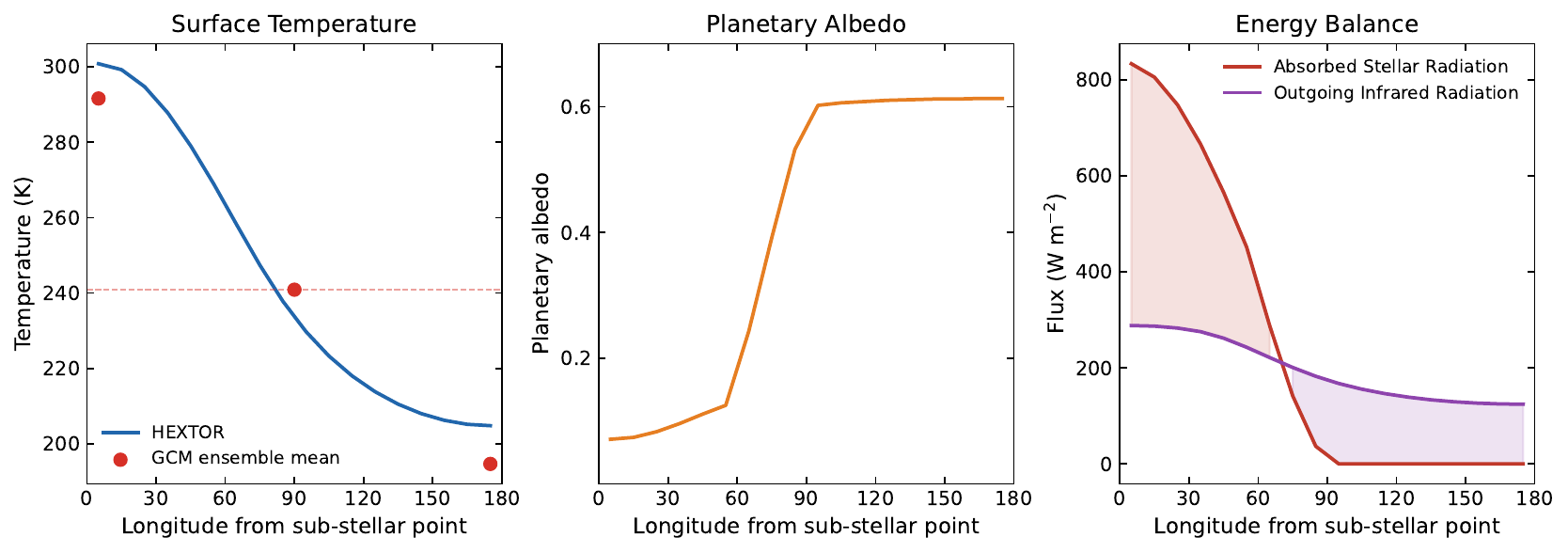}
    \caption{HEXTOR calculations of the THAI Hab 1 case for TRAPPIST-1 e (blue curve), calibrated to the THAI GCM ensemble mean values for minimum, average, and maximum surface temperature (red circles). Planetary albedo is highest on the icy nightside, where net radiative cooling occurs.\vspace{0.20cm}}
    \label{fig:Hab1}
\end{figure*}

Planets in synchronous rotation around low-mass host stars are difficult to simulate with a latitudinal EBM, because the bulk energy transport on such a planet is from the sub-stellar dayside of the planet to the anti-stellar nightside. One option is to transform the coordinate axis of the EBM so that it represents longitude instead of latitude, with the instellation function also modified so stellar flux is only incident on one hemisphere \citep{fortney2010transmission,koll2015deciphering,checlair2017no}. This tidally-locked coordinate transformation was implemented in HEXTOR \citep{haqq2022energy} by assigning the variable $x$ in Eq. (\ref{eq:EBM}) to represent the cosine of the tidally locked longitude (ranging from the sub-stellar point at 0\degree to 180\degree, with the terminator at 90\degree). The function for instellation varies with zenith angle, $\phi_z$, so that instellation is $S\cos \phi_z$ on the dayside ($\phi_z \leq 90\degree$) and 0 on the nightside ($\phi_z > 90\degree$). The present version of HEXTOR includes corrections to the tidally-locked coordinate transformation, which are discussed in Appendix \ref{ap:updates2}.

Reference cases for the atmospheres of planets around low-mass stars were developed by the TRAPPIST-1 Habitable Atmosphere Intercomparison (THAI) project, which were used to compare several general circulation models under common configurations relevant to TRAPPIST-1 e \citep{fauchez2020trappist,turbet2022trappist,sergeev2022trappist,fauchez2022trappist}. These reference cases include two benchmark configurations with dry land surfaces (Ben 1 and Ben 2) and two experiment cases with ocean surfaces and moist atmospheres (Hab 1 and Hab 2). The Ben 1 and Hab 1 cases use a 1\,bar N$_2$ + 400\,ppm CO$_2$ atmosphere, while the Ben 2 and Hab 2 cases use a 1\,bar CO$_2$ atmosphere. The radiative transfer lookup tables for HEXTOR were generated specifically for 1\,bar N$_2$ atmospheres with an additional contribution from CO$_2$ partial pressure \rev{(ranging from 10$^{-4}$\,bar to 10\,bar)}, with  an ocean surface and water vapor present in the atmosphere, so the Hab 1 experiment is the only THAI case that is consistent with the physical limitations of the model. (Note that the previous version of HEXTOR was used in a conceptual exploration of all four THAI cases \citep{haqq2022energy}, but the Hab 1 case remains the one aligned with the model physics.) 

The HEXTOR configuration for the THAI Hab 1 case follows the parameters of the THAI protocol \citep{fauchez2020trappist} to the greatest extent feasible with an EBM. The radiative transfer lookup table uses a 2600\,K host star, with insolation $S_0=900$\,W\,m$^{-2}$ and orbital period of 6.1\,d. Atmospheric composition is 1\,bar N$_2$ with an additional 400\,ppm CO$_2$, with complete ocean coverage (no land) at all latitudinal bands. Surface albedo values are fixed for ocean at 0.06 and ice at 0.25, and the heat capacity is $C=4\times10^{6}$\,J\,m$^{-2}$K$^{-1}$ for both ocean and ice. \rev{Because instellation has no time dependence for a synchronous rotator at zero obliquity, the steady state is insensitive to the value of $C$, which only sets the spin-up timescale: repeating the calibrated case below with heat capacities from $5.25\times10^{6}$ to $2.1\times10^{8}$\,J\,m$^{-2}$K$^{-1}$ (the values used for land and ocean in the Earth configuration) changes the equilibrium temperatures by less than 0.2\,K, while the convergence time grows from ${\sim}50$ to ${\sim}2000$ model years. The adopted value was chosen for rapid convergence across large parameter sweeps.} The finite differencing timestep is reduced to $\Delta t=0.375$\,hr, and obliquity is set to zero. With the tidally locked coordinate transformation, this EBM configuration can be used to explore possible climates for TRAPPIST-1 e.

Figure \ref{fig:Hab1} shows HEXTOR calculations of the THAI Hab 1 case, which have been calibrated to the ensemble mean values for minimum, average, and maximum surface temperature from the four general circulation models participating in THAI \citep{sergeev2022trappist}. The calibration was performed by running a large number of EBM simulations that varied the value of the diffusion parameter $D$ as well as the cloud infrared parameter (a reduction to $F$), with the goal of matching the ensemble global average temperature and the day-night temperature contrast. This gave calibrated values of \rev{$D=3.10$}\,W\,m$^{-2}$\,K$^{-1}$ and a cloud infrared reduction to $F$ of \rev{$-35$}\,W\,m$^{-2}$. The resulting calibrated calculation gives a global mean of \rev{240.9}\,K, a near match (off by \rev{less than 0.1\,K}) to the THAI ensemble. The maximum surface temperature at the sub-stellar point is \rev{300.8}\,K (\rev{9.2}\,K higher than the THAI ensemble) and the minimum surface temperature at the anti-stellar point is 204.8\,K (10.1\,K higher than the THAI ensemble). Planetary albedo is highest on the anti-stellar hemisphere where ice cover dominates. (Note that these calculations do not include a sub-stellar cloud deck with high albedo as predicted by \cite{yang2013stabilizing}, as EBMs such as HEXTOR lack explicit representation of convection or clouds.) The energy balance shows all stellar radiation absorbed in the sub-stellar hemisphere, with outgoing infrared radiation dominating the anti-stellar hemisphere.

\rev{The residual calibration errors in the minimum ($+10.1$\,K) and maximum ($+9.2$\,K) surface temperatures represent a structural limitation of the one-dimensional diffusive model, which cannot simultaneously match the global mean, the sub-stellar maximum, and the anti-stellar minimum with a single constant $D$; these residuals persist nearly unchanged across all of the alternative calibrations described below. Absolute temperatures from the model should therefore be regarded as uncertain at the ${\sim}10$\,K level, and the conclusions of this study are drawn from the classification of ice states rather than from absolute temperature values.}

\rev{The calibrated diffusion parameter is ${\sim}5$ times larger than the Earth-tuned value of Section 2, and two physical interpretations are possible: efficient day--night energy transport by the global-scale circulation of synchronous rotators, or a compensation for the missing reflective sub-stellar cloud deck of \cite{yang2013stabilizing}, which would otherwise reduce the energy contrast between the two hemispheres. To distinguish between these, the model was extended with an option for a prescribed dayside cloud deck: a constant cloud albedo $a_c$ is blended into the surface albedo with cloud fraction $f_c$, which affects only the dayside because the nightside receives no stellar flux. Adding a moderate deck ($f_c=0.5$, $a_c=0.5$) to the calibrated model cools the global mean from 240.9 to 235.8\,K and moves the dayside ice line from 60\degree\ to 54\degree\ from the sub-stellar point. Recalibrating $D$ and the cloud infrared parameter with the deck active yields $D=2.90$\,W\,m$^{-2}$\,K$^{-1}$ and a cloud infrared reduction of $-30$\,W\,m$^{-2}$ for the moderate deck, and $D=2.50$\,W\,m$^{-2}$\,K$^{-1}$ with $-4$\,W\,m$^{-2}$ for a strong deck ($f_c=0.8$, $a_c=0.6$). The cloud infrared parameter thus absorbs most of the missing shortwave cloud forcing (shrinking from $-35$ to $-4$\,W\,m$^{-2}$), while the calibrated $D$ decreases only modestly and remains ${\sim}4.3$ times the Earth-tuned value. The large diffusion parameter is a robust feature of day--night heat transport on synchronous rotators rather than a proxy for missing cloud albedo. Both recalibrated cloud-deck variants are carried through the full parameter space analysis below.}

\begin{figure*}[tbp]
    \centering
    \includegraphics[width=\textwidth]{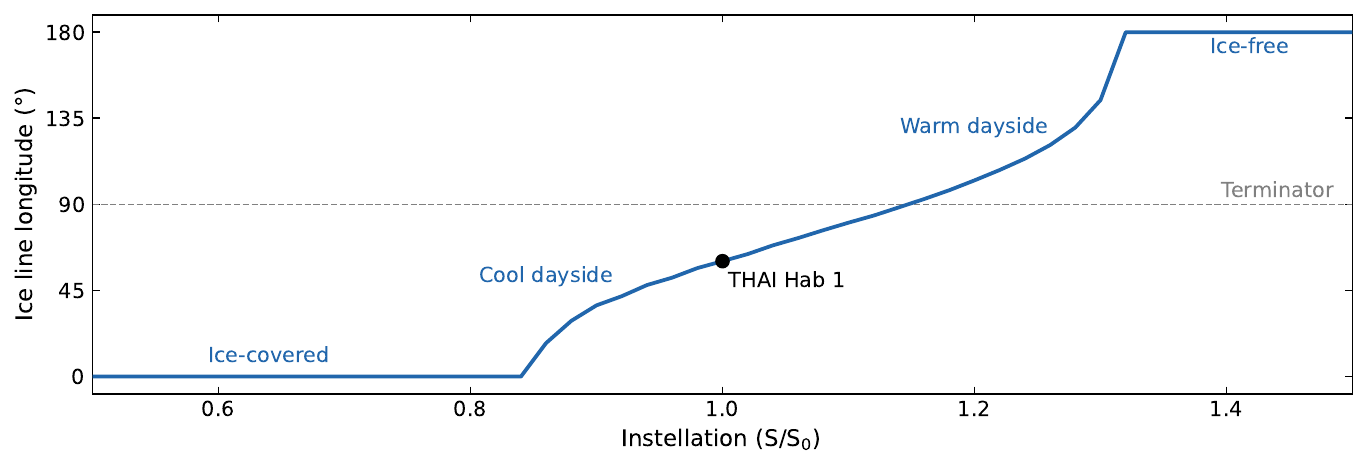}
    \caption{Ice states for HEXTOR calculations based on the THAI Hab 1 configuration, which show smooth transitions from ice-covered to cool dayside to warm dayside to ice-free states as instellation increases. These results do not depend on the choice of initialization temperature.}
    \label{fig:THAIice}
\end{figure*}
\begin{figure*}[tbp]
    \centering
    \includegraphics[width=\textwidth]{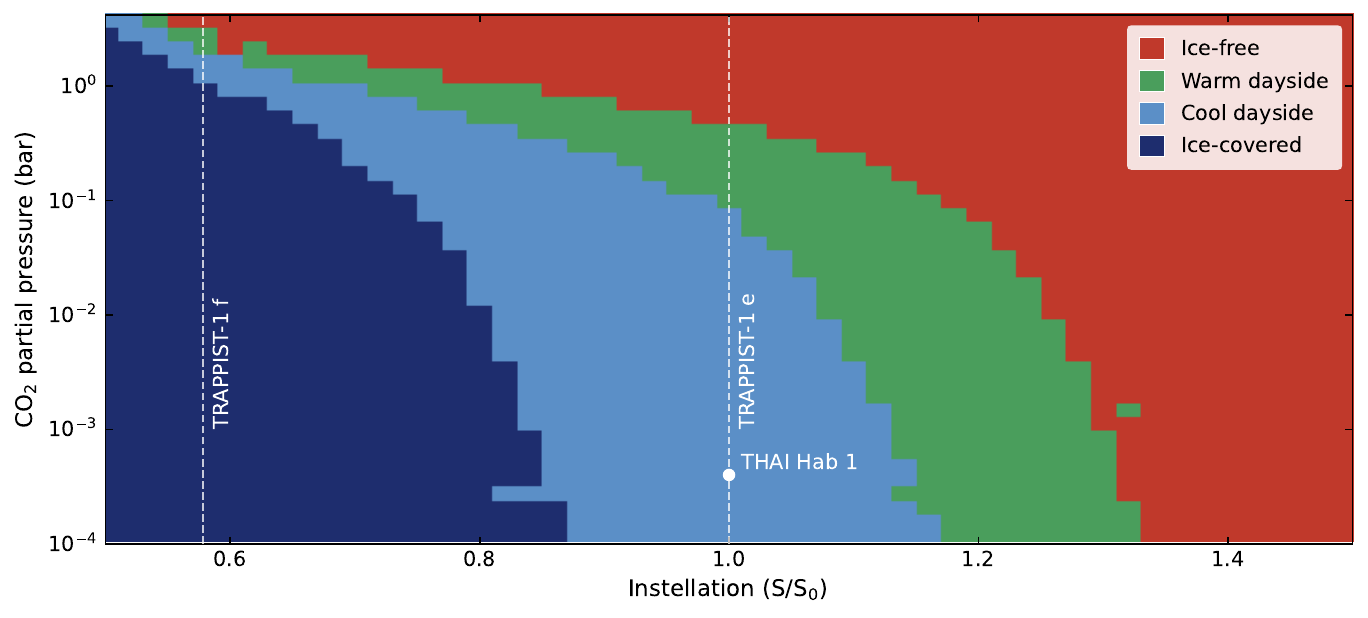}
    \caption{Ice states for HEXTOR calculations based on the THAI Hab 1 configuration, across a parameter space of instellation and CO$_2$ partial pressure.\vspace{0.20cm}}
    \label{fig:THAIphase}
\end{figure*}

This calibrated Hab 1 case is next used to explore the different ice coverage phases as a function of instellation. Planets in synchronous rotation around low-mass stars are not expected to show hysteresis \citep{checlair2017no,checlair2019no}, as occurs for Earth-like climates (Fig. \ref{fig:phases}, top row), and instead should produce the same climate state regardless of initial conditions. The calculations shown in Figure \ref{fig:THAIice} find that the present version of HEXTOR gives the same transitions from an ice-covered to an ice-free state for different values of initial temperature (trials included 288\,K, 273\,K, and 233\,K). The previous version of HEXTOR \citep{haqq2022energy} showed limited hysteresis when attempting to simulate the THAI cases, but this behavior was correctly identified as ``likely a model artifact, rather than a physical result'' that vanished with the corrections to the model (Appendix \ref{ap:updates} and \ref{ap:updates2}). The ``warm dayside'' phase applies when the longitudinal ice line is restricted to the anti-stellar hemisphere, while the ``cool dayside'' phase applies when the longitudinal ice line crosses the terminator and extends into the sub-stellar hemisphere. These phases are continuous transitions, rather than abrupt changes, but they can be instructive in using an EBM to characterize climate across a large parameter space.

A complete phase space for this calibrated model is shown in Figure \ref{fig:THAIphase}, which summarizes the results of \rev{4788} HEXTOR simulations with both instellation and CO$_2$ partial pressure varied across the range of the model's capabilities. \rev{In these calculations the N$_2$ inventory is fixed at 1\,bar with CO$_2$ partial pressure added on top (so the 0.5\,bar CO$_2$ case has a total surface pressure of 1.5\,bar), consistent with the construction of the lookup table, and $D$ is held fixed at its calibrated value (the sensitivity of the results to this assumption is quantified below).} This places TRAPPIST-1 e in a cool dayside state up to a CO$_2$ partial pressure of about \rev{$7\times10^{-2}$\,bar}, with possible transitions to a warm dayside or ice-free state at higher values of CO$_2$. The CO$_2$ partial pressure for the ice-free transition (\rev{$\sim4\times10^{-1}$\,bar}) is \rev{about a factor of two above} the ice-free transition found in simulations with the ExoCAM general circulation model of TRAPPIST-1 e \citep[][Fig. 3]{wolf2018erratum}\rev{, a difference comparable to the sensitivity of this threshold to the cloud parameterization (Table \ref{tab:sensitivity})}. It is worth noting that TRAPPIST-1 e remains in a cool dayside state at the lowest simulated CO$_2$ values in the EBM and does not enter an ice-covered state. \rev{This cool dayside (or ``eyeball'') regime at low CO$_2$ is qualitatively consistent with three-dimensional simulations of TRAPPIST-1 e \citep{turbet2018modeling,wolf2018erratum,sergeev2022trappist}; the contribution of the EBM is to map where this regime begins and ends across a parameter space that remains impractical to survey with general circulation models.}

The phase space also shows TRAPPIST-1 f in an ice-covered state up to a CO$_2$ partial pressure of about \rev{1.2\,bar}. \rev{This is consistent with} simulations with the ExoCAM general circulation model of TRAPPIST-1 f\rev{, which} found ice-covered conditions persisted even at 1\,bar CO$_2$, although a transition to ice-free conditions occurred in ExoCAM by 2\,bar CO$_2$\rev{, compared to ${\sim}3$\,bar in the EBM. This factor of ${\sim}1.5$ difference in the ice-free transition is well within the sensitivity of this threshold to the cloud parameterization, which spans 1.5\,bar to beyond the upper limit of the lookup table across the sensitivity ensemble described below (Table \ref{tab:sensitivity})}. (The planet TRAPPIST-1 d is at $S/S_0=1.73$, within the ice-free regime but beyond the physical limits of the model.) These calculations demonstrate the capability of a calibrated EBM using tidally locked coordinates to explore a large parameter space, which can be compared with more computationally expensive models and also guide such models toward the regions of parameter space worth exploring in greater detail.

\begin{figure*}[tbp]
    \centering
    \includegraphics[width=\textwidth]{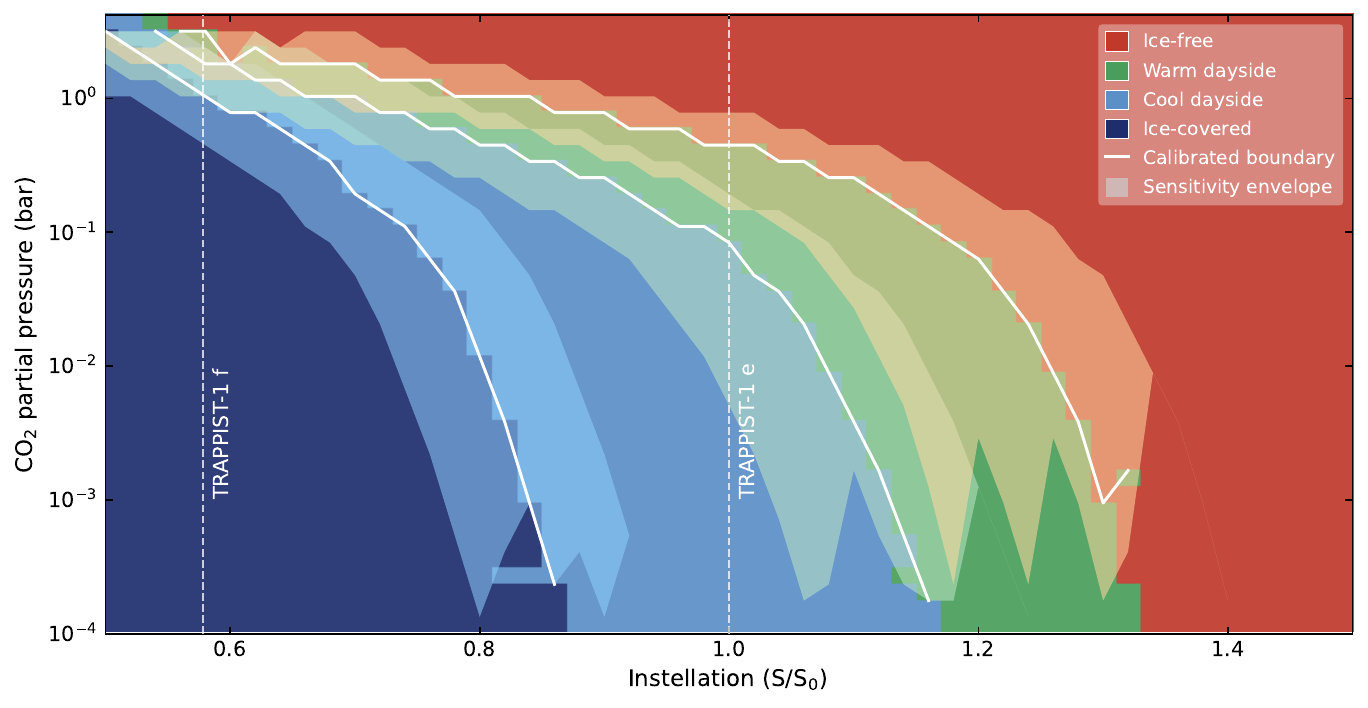}
    \caption{\rev{Sensitivity of the phase boundaries in Figure \ref{fig:THAIphase} to the calibrated model parameters. The filled background shows the calibrated model, white curves show its three phase boundaries, and the shaded envelopes show the range spanned by those boundaries across the seven-member sensitivity ensemble ($D\times0.8$, $D\times1.2$, cloud infrared parameter of $-17.5$ and $-52.5$\,W\,m$^{-2}$, the two recalibrated dayside cloud-deck variants, and the calibrated baseline).}\vspace{0.20cm}}
    \label{fig:sensitivity}
\end{figure*}
\begin{table*}[t]
    \centering
    \begin{tabular}{lcccc}
        \hline
         & \multicolumn{2}{c}{TRAPPIST-1 e} & \multicolumn{2}{c}{TRAPPIST-1 f} \\
        Sensitivity experiment & cool$\rightarrow$warm & ice-free & leaves ice-covered & ice-free \\
        \hline
        Calibrated                    & 0.074 & 0.445 & 1.17 & 3.12 \\
        $D\times0.8$                  & 0.067 & 0.559 & 0.72 & 3.68 \\
        $D\times1.2$                  & 0.079 & 0.369 & 1.39 & 2.81 \\
        Cloud IR $-17.5$\,W\,m$^{-2}$ & 0.010 & 0.232 & 0.44 & 1.72 \\
        Cloud IR $-52.5$\,W\,m$^{-2}$ & 0.208 & 0.751 & 2.55 & $>$10 \\
        Cloud deck (moderate)         & 0.064 & 0.413 & 0.81 & 2.80 \\
        Cloud deck (strong)           & 0.005 & 0.166 & 0.53 & 1.50 \\
        \hline
    \end{tabular}
    \caption{\rev{Sensitivity experiments showing CO$_2$ partial pressure thresholds (bar) for climate state transitions at the instellations of TRAPPIST-1 e ($S/S_0=1.0$) and TRAPPIST-1 f ($S/S_0=0.578$), refined by bisection to ${\sim}5$\% precision.}\vspace{0.5cm}}
    \label{tab:sensitivity}
\end{table*}

\rev{Because the model is calibrated against a single reference case and those parameters are then held fixed across thousands of simulations, the sensitivity of the phase boundaries to the calibration was quantified by repeating the full phase-space calculation for six perturbed model versions: the diffusion parameter varied by $\pm$20\%, the cloud infrared parameter halved ($-17.5$\,W\,m$^{-2}$) and increased by half ($-52.5$\,W\,m$^{-2}$), and the two recalibrated dayside cloud-deck variants described above. Figure \ref{fig:sensitivity} shows the envelope of the resulting phase boundaries, and Table \ref{tab:sensitivity} lists the transition thresholds at the instellations of TRAPPIST-1 e and f. The classification of TRAPPIST-1 e remains robust: the planet resides in a cool dayside state at the lowest CO$_2$ values in all seven sensitivity experiments (with the dayside ice line ranging from 40\degree\ to 71\degree\ from the sub-stellar point), never entering an ice-covered state, and the CO$_2$ threshold for the transition out of the cool dayside state is nearly unaffected by $\pm$20\% variations in $D$ (0.067--0.079\,bar). The dominant uncertainty is instead the treatment of cloud radiative effects: across the ensemble, the TRAPPIST-1 e ice-free threshold spans 0.17--0.75\,bar and the TRAPPIST-1 f ice-free threshold spans 1.5\,bar to beyond the table limit ($>$10\,bar). The qualitative structure of Figure \ref{fig:THAIphase} is therefore robust to the calibration choices, while the precise CO$_2$ values of the ice-free transitions carry a factor of ${\sim}2$--3 uncertainty dominated by the cloud parameterization.}

\vspace{0.20cm}\section{Discussion \& Conclusion}

This model development study has featured the capability of a one-dimensional energy balance model to provide physical intuition about the climates of planets in synchronous rotation around low-mass stars. If observations point toward a non-zero CO$_2$ abundance on TRAPPIST-1 e, then these calculations would suggest that TRAPPIST-1 e is likely to be in a cool dayside state---unless CO$_2$ partial pressure is large (\rev{$\gtrsim0.07$\,bar, with the transition to a fully ice-free state at ${\sim}0.4$\,bar}). Likewise, if observations suggest an atmosphere with CO$_2$ on TRAPPIST-1 f, then these calculations would suggest that TRAPPIST-1 f is likely to be in an ice-covered state---again, unless partial pressure approaches \rev{${\sim}1.2$\,bar} CO$_2$. These EBM calculations provide an immediately accessible way to interpret observations of exoplanet atmospheres, which may be more physically meaningful than using the raw value of stellar flux or a derived ``equilibrium temperature'' to infer climate properties. 

\rev{Several caveats accompany these predictions, beyond the parameter sensitivity quantified above. First, the THAI Hab 1 calibration targets are derived from general circulation models with immobile slab oceans; fully coupled simulations show that dynamic ocean heat transport can carry open water well onto the nightside of synchronous rotators \citep{hu2014role}, which would substantially modify the day--night temperature contrast to which $D$ is calibrated. Second, the model assumes a flat surface, whereas recent three-dimensional simulations show that topographic relief alone can lower the critical CO$_2$ pressure for nightside deglaciation by a factor of several through topographically forced stationary waves \citep{chen2026topography}, making surface relief a first-order control on the ice-covered boundary \citep[see also][]{guimond2026water}. Third, the absence of hysteresis found here (Fig. \ref{fig:THAIice}) applies to strictly synchronized planets: planet--planet interactions in compact systems such as TRAPPIST-1 can drive libration of the sub-stellar point, and ice formed during a libration excursion can persist through the ice-albedo feedback, reintroducing a form of climate hysteresis \citep{chen2023sporadic}. Finally, thermal atmospheric tides can drive planets into asynchronous or higher-order spin states such as a 3:2 resonance \citep{leconte2015asynchronous}; such a planet would have no permanent dayside or nightside, and its slowly migrating sub-stellar point would produce ice coverage and temperature distributions qualitatively different from those modeled here.}

In such a scenario where CO$_2$ is observed on TRAPPIST-1 e or f, the next step would be to investigate the most relevant region of a parameter space for atmospheric composition using a three-dimensional general circulation model. Such models can provide more detailed assessments of possible climate states, but they are also computationally expensive. The results of any observationally-informed three-dimensional model calculations could then be used further as calibration for the EBM, thereby allowing the one-dimensional model to perform a large number of calculations across the region of interest. This iterative approach between three-dimensional models and one-dimensional models can help to constrain planetary properties and update such predictions in light of new observational data.

Fully realizing this iterative model framework will also require extending the radiative transfer capabilities of HEXTOR to a broader set of atmospheres. The calculations in this paper assumed an atmospheric composition limited to N$_2$, CO$_2$, and H$_2$O, due to the data available in the radiative transfer lookup tables, but these are far from the only gaseous species present in exoplanet atmospheres. One way to extend this capability to other atmospheric compositions is to re-derive new lookup tables (or equivalently, develop new high-order polynomial fits) that can be selected as needed based on the specific atmosphere being investigated. However, this approach is inherently limiting, and a better option would enable HEXTOR to investigate atmospheres with arbitrary compositions without needing to generate a new lookup table. 

The best method would be to couple the EBM with a radiative transfer column model at each latitudinal band, in order to calculate the surface radiative fluxes explicitly during runtime. An approach like this would have been infeasible when Budyko-Sellers models were first developed, but recent advances in parallel computing make such problems tractable. Further development of HEXTOR will allow the model to become a robust tool for mapping out the climate states of terrestrial planets in order to guide observations for exoplanet characterization and the search for life.

\section*{Acknowledgments}
Thanks to Thomas Fauchez, Ravi Kopparapu, and Eric Wolf for numerous conversations prior to and during this study. The author gratefully acknowledges support from the NASA Habitable Worlds program under grant 80NSSC24K1896. Any opinions, findings, and conclusions or recommendations expressed in this material are those of the author and do not necessarily reflect the views of any employer or NASA.
\\\\
\textit{Software:} HEXTOR \citep[v4.2.1][]{jacob_haqq_misra_2026_20074009}, Matplotlib \citep{Hunter:2007}, Claude Code \citep[Opus 4.8,][]{anthropic2026opus48} 
\vspace{0.15cm}

\bibliographystyle{mnras}

\bibliography{main}

\begin{appendix}

\section{Corrections to the Lookup Table}\label{ap:updates}

The HEXTOR lookup table contains 1748 values of $F$ and 34,960 values of $\alpha$ that were calculated with a radiative-convective equilibrium climate model \citep{kopparapu2013habitable} over a parameter space of CO$_2$ partial pressure \rev{($10^{-4}$\,bar to 10\,bar of CO$_2$ added to the fixed 1\,bar of N$_2$, corresponding to total surface pressures of 1--11\,bar)}, surface temperature (190\,K to  370\,K), zenith angle (0\degree to 90\degree), and surface albedo (0.2 to 1). The radiative-convective calculations assume a surface ocean, with atmospheric water vapor content based on the Clausius-Clapeyron relation. The model interpolates between these lookup table values at runtime using a nearest-neighbor search. \rev{Outside the tabulated ranges of CO$_2$ and surface albedo, the interpolation clamps to the nearest table boundary: CO$_2$ partial pressures below the table minimum of $10^{-4}$\,bar reuse the lowest tabulated value, so Figures \ref{fig:THAIphase} and \ref{fig:sensitivity} are truncated at this floor rather than extending to lower CO$_2$ where the radiative transfer would simply repeat the table minimum; and surface albedo values below 0.2 act as an effective shortwave surface albedo of 0.2 (which applies to the nominal ocean albedo of 0.06 used in this work).}

The original lookup table \citep{haqq2022energy} contained errors in the interpolation functions for $\alpha$ and $F$ that were causing spurious behavior at certain regions of the parameter space. These errors were relatively small but pronounced at certain zenith angles and high carbon dioxide partial pressure, as shown in Figure \ref{fig:errors}. These improvements only led to a minor difference (less than 0.1\,K) in results for present-Earth conditions.

The calculation for $\alpha$ was also incorrectly using the cosine of the zenith angle, rather than the zenith angle itself, which led to an underestimation of $\alpha$. In the original lookup table \citep{haqq2022energy}, this was compensated for by reducing the value of $F$ by 8.0\,W\,m$^{-2}$ with a physical justification of representing clouds. This has now been fixed in the present version of HEXTOR, which now uses a cloud infrared parameter of \rev{6.7}\,W\,m$^{-2}$ to produce a pre-industrial Earth state.

\rev{A further error was identified in the CO$_2$ coordinate of the lookup tables during revision. The tabulated CO$_2$ coordinate is the mixing ratio corresponding to each 1\,bar N$_2$ + CO$_2$ atmosphere, but the stored values had been computed with an incorrect background pressure of $\pi$\,bar for the M-dwarf table and $\pi/2$\,bar for the solar table (an error likely related to the instellation factor discussed in Appendix \ref{ap:updates2}), rather than the correct 1\,bar of N$_2$. Because the model queries this coordinate at runtime, each lookup effectively retrieved the radiative transfer of an atmosphere with roughly $\pi$ (or $\pi/2$) times the intended CO$_2$ inventory. The coordinate has been rewritten as the true mixing ratio, the model now performs lookups in mixing-ratio space, and all tuning and calibration in this paper was repeated with the corrected tables: the pre-industrial Earth state requires the cloud infrared parameter of 6.7\,W\,m$^{-2}$ quoted above, and the THAI Hab 1 calibration of Section 3 yields $D=3.10$\,W\,m$^{-2}$\,K$^{-1}$ with a cloud infrared reduction of $-35$\,W\,m$^{-2}$. The corrections shift the CO$_2$ values of climate-state transitions upward by a factor of ${\sim}\pi$ relative to the uncorrected model but leave the qualitative structure of all results unchanged.}

\onecolumngrid
\vspace{\intextsep}
\noindent\includegraphics[width=\textwidth]{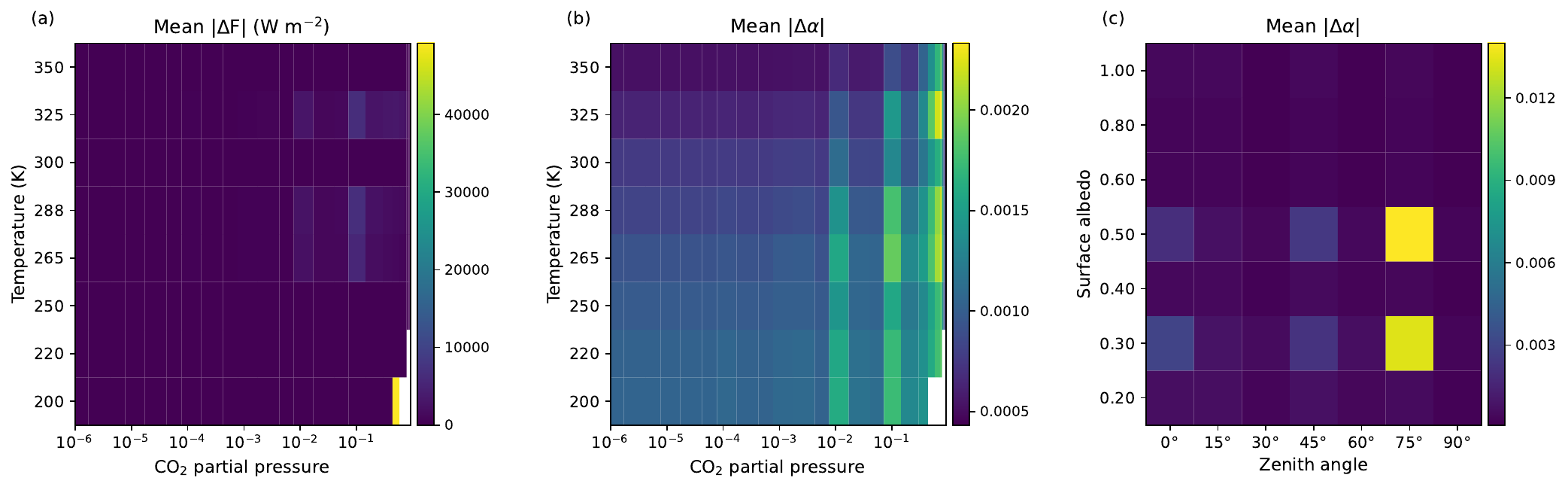}
\figcaption{Mean differences between the previous version of the HEXTOR radiative transfer lookup table \citep{haqq2022energy} and the improved version described in this paper. The mean outgoing infrared radiation (panel a) shows the largest errors at high CO$_2$ partial pressure and low temperature. The mean planetary albedo shows discrete regions of large error at certain high values of CO$_2$ partial pressure (panel b) as well as at specific values of zenith angle (panel c). \rev{These panels isolate the corrections to the interpolation functions and zenith-angle treatment; because both versions are evaluated on the same lookup table, the separate correction to the CO$_2$ mixing-ratio coordinate (described above) cancels in this difference and is not shown here.}\label{fig:errors}}
\vspace{\intextsep}

\section{Corrections to the Coordinate Transformation}\label{ap:updates2}

A separate lookup table for HEXTOR was constructed by \cite{haqq2022energy} with an M-dwarf (2600\,K) host star, using the same radiative-convective model \cite{kopparapu2013habitable} and exploring the same parameter space \rev{(the corrected CO$_2$ coordinate described in Appendix \ref{ap:updates} applies to both tables, which share the same CO$_2$ partial pressure grid of $10^{-4}$\,bar to 10\,bar)}. The present version of HEXTOR has implemented a power-based extrapolation function for end cases of high and low temperatures that exceed the bounds of the lookup table. Such cases did not occur during the Earth tuning but did occur with some tidally-locked test cases. The values of this extrapolation are only reasonable up to $\sim$400\,K, beyond which the results lose physical meaning. The results shown in this paper are restricted to cases that remain below this threshold.

The values of zenith angle and instellation were also incorrectly calculated in the original HEXTOR implementation \citep{haqq2022energy}. In the present version, the correct value for zenith angle is used and an extra factor of $\pi^{-1}$ has been removed from the functional form of $S$. These corrections allow the model to better match the results of more computationally expensive models, as discussed in the main text.

\end{appendix}

\end{document}